\date{June 20th}

\documentclass[reqno]{amsart}
\usepackage{amsmath,amsfonts,amssymb,amsthm}
\usepackage{graphicx,epsfig} 

\numberwithin{equation}{section}
\newtheorem{thm}{Theorem}
\newtheorem{lemma}{Lemma}[section]

\newcommand{\remark}{{\it Remark: \,\,}}

\newcommand{\dist}{{\, \rm dist}}

\newcommand{\Z}{\mathbb Z}

\newcommand{\T}{\mathbb T}

\newcommand{\e}{\mathrm e}

\newcommand{\R}{\mathbb R}
\newcommand{\N}{\mathbb N}

\newcommand{\E}{\mathbb E}

\theoremstyle{definition}
\newtheorem{defn}{Definition}

\begin{document}
\title[Lifshitz tails in the $3D$ Anderson model]
{Lifshitz tails and Localization in $3D$ Anderson model}

\author[A. Elgart]{Alexander Elgart}
\address{A. Elgart \\ Ben Gurion University}
\email{aelgart@vt.edu}

\begin{abstract}  Consider the $3D$
 Anderson model with a zero mean and bounded i.i.d. random
potential. Let $\lambda$ be the coupling constant measuring the
strength of the disorder, and $\sigma(E)$  the self energy of the
model at energy $E$. For any $\epsilon>0$ and sufficiently small
$\lambda$, we derive almost sure localization in the band $E\le
-\sigma(0)-\lambda^{4-\epsilon}$. In this energy region, we show
that the typical correlation length $\xi_E$ behaves roughly as
$O((|E|-\sigma(E))^{-1/2})$, completing the argument, outlined in
the unpublished work of T. Spencer \cite{Spencer}.
\end{abstract}
\maketitle


\section{Introduction, main result and steps of the
proof}\label{sec:intro}
\subsection{Introduction}
In this paper we want to carry out the program, sketched in the
unpublished preprint of T. Spencer \cite{Spencer}, regarding the
localization for the $3D$ Anderson model in the so-called Lifshitz
tails regime.

The {\em Anderson} operator $H^\lambda_\omega$ on the lattice $\Z^3$
acts on the vector $\psi\in l^2(\Z^3)$ as:
\begin{equation}\label{eq:model}
(H^\lambda_\omega \psi)(n):=-\frac{1}{2}(\Delta \psi)(n)+\lambda
V_\omega(n) \psi(n)\,,
\end{equation}
 where $\Delta$ denotes the discrete Laplace
operator,
$$(\Delta \psi)(n) \ = \  \sum_{e\in \Z^3,\ |e|=1}\psi(n+e)\ - \ 6 \psi(n) \,.$$ We
will assume throughout this paper
\begin{enumerate}
\item[A1] The values of the random potential $V_\omega(\cdot)$ are
independent, identically distributed variables, with even, compactly
supported, and bounded probability density $\rho$.

\item[A2] For any $m\in\N$, $\E (V^{2m}_\omega(x))\le c$ with some constant $c$,
and $\E\, (V_\omega^2(x))=1$.
\end{enumerate}
Let $e(p)$ denote the dispersion law, associated with the Fourier
transform of the Laplacian, $({\mathcal F} \Delta f)(p)= -2 e(p)\hat
f(p)$, where
$$\hat f( p):=({\mathcal F}f)( p):=\sum_{n\in\Z^3}e^{-i2\pi p\cdot
 n}f(n)\,,\quad   p \in\T^3:=[-1/2,1/2]^3\,,$$ with its inverse
$$\check g(  n )=\int_{\T^3}d^3 p \,e^{i2\pi  p \cdot
 n }f(  p )\,.$$
 One then computes
\begin{equation}
e( p)=2\sum_{\alpha=1}^3\sin^2(\pi p\cdot e_\alpha)\,,
\end{equation}
where  $e_\alpha$ is a unit vector in the $\alpha$ direction. The
spectrum of the unperturbed operator $H^0_\omega$ is absolutely
continuous and consists of the interval $[0,6]$.

In what follows we will denote by $A(x,y)$ the kernel of the linear
operator $A$ acting on $l^2(\Z^3)$ (that is $A(x,y)=(\delta_y,
A\delta_x)$, where $\delta_x$ is an indicator function of the site
$x\in \Z^3$, and $(\cdot,\cdot)$ denotes the inner product of
$l^2(\Z^3)$). We will use the concise notation $\int$ in place of
$\int_{(\T^3)^{k}}$ whenever it is clear from the context that each
of the $k$ variables of integration is integrated out over a torus
$\T^3$.

 We will investigate the
 properties of $H_\omega^\lambda$ for a typical
configuration $\omega$ in a weak disorder regime, namely at the
energy range \[\left[\lambda a\,,\,-\lambda^2\int_{\T^3}
\frac{d^3p}{e(p)} -\lambda^{4-\epsilon}\right]\,,\] with
$a=\min\{x:x\in{\rm support}\ \rho\}$ for any $\epsilon>0$ and
$\lambda>0$ being sufficiently small\footnote{$\lambda
a=\inf(E:E\in\sigma(H^\lambda_\omega))$ almost surely, see e.g.
\cite{Stollmann} for details.}. Most of the mathematical results on
localization for operators with random potential in dimensions $d
>1$ have been derived using the multi-scale analysis introduced by
Fr\"ohlich and Spencer \cite{FS} and by the fractional moment method
of Aizenman and Molchanov \cite{AM} (we are going to use the latter
approach). By now there exists extensive general literature on the
Anderson localization problem, see for example \cite{Stollmann} and
references therein.

The quantity of the most interest is the typical asymptotic behavior
of the so called Green function (also known as the two point
correlation function, the propagator)
\[R(x,y) \ = \ (H^\lambda_\omega+E+i\eta)^{-1}(x,y)\]
in the limit $\eta\searrow0$. It plays a crucial role in
determining, for instance, the conductivity properties of the
physical sample (whether it is an insulator or a conductor at a
given energy band). On a mathematical level, investigation of the
propagator can yield an insight on the typical spectrum of
$H_\omega^\lambda$ at the vicinity of $-E$. The Anderson model
(\ref{eq:model}) is characterized by the following dichotomy,
\cite{ASFH}: Either the typical Green function $R(x,y)$ decays at
least exponentially fast when $|x-y|\rightarrow\infty$, or it cannot
decay faster than $|x-y|^{-6}$ in a three dimensional case. The
former behavior is called {\em localization} and necessitate that
the spectrum is pure point almost surely in the vicinity of $-E$,
\cite{AM}.  The localization region is naturally characterized by
the so called {\em correlation length } $\xi_E$, that is a typical
length scale $|x-y|$ at which $R(x,y)$ starts to decay at least
exponentially fast. We stress here the energy dependance of the
correlation length, for it is going to play a role in our analysis.
The consensus among the condensed matter physicists is that in $3D$,
in the weak disorder regime, there should be a spectral transition
from point spectrum to continuous one. This phenomenon is known as
the Anderson transition. The proof of such an actuality presents a
great challenge in this subject. The region of the possible spectral
transition is called the {\em mobility edge}. One can get a certain
indication of the existence of the delocalized regime for the
Anderson model by showing that there occurs a threshold energy $E_0$
(which presumably coincides with the mobility edge of the problem)
such that $\xi_E$ diverges as $E$ approaches $E_0$. The present work
can be seen as a step in this direction.

The occurrence of localization at energies near the band edges at
weak disorder is related to the rarefaction of low eigenvalues, and
was already discussed in the physical literature by I.~M. Lifshitz
in 1964, see Section $3$ in \cite{Lifshitz1},  and \cite{Lifshitz2}.
As far as the rigorous results are concerned, let us only mention
the three closely related works: M.~Aizenman \cite{A} showed that
the spectrum of $H_\omega^\lambda$ consists (almost surely) of the
localized eigenvalues at the energy range $[\lambda a, \lambda a
+\lambda^{\alpha}]$, with  $\alpha=5/4$. This result was later
improved by W-M.~Wang \cite{Wang} ($\alpha=1$), whose result was in
turn enhanced by F.~Klopp \cite{Klopp}, who extended the region of
localization all the way up to the (negative) energies of order
$\lambda^{1+1/6}$ in 3$D$. In this work we push the upper bound of
the localization region further up, to the value $-C\lambda^2$, and
examine the behavior of the correlation length as a function of
energy.
\subsection{Results}
In order to formulate our main technical accomplishment we need to
introduce some further notation. The self energy  term $\sigma(E)$,
associated with  $H^\lambda_\omega$, is given by the solution of the
self-consistent equation\footnote{Note that since we will be
interested at negative energies, $E$ in (\ref{eq:self}) is assumed
to be positive.}
\begin{equation}\label{eq:self}
\sigma(E)=\lambda^2\int_{\T^3} \frac{d^3p}{e(p)+E-\sigma(E)}\,.
\end{equation}
It is easy to check that $\sigma(E)$ is positive and uniformly
bounded by $C \lambda^2$ with some constant $C$, provided that
\[E\ > \ E_0:= \lambda^2\int \frac{d^3p}{e(p)}\]  and $E^*:=E-\sigma(E)>0$ for
such values of $E$ (the  relevant properties of the solution of
(\ref{eq:self}) are collected in Appendix \ref{sec:appendI}.
Moreover, if
\begin{equation}\label{eq:range}E\ge E_\epsilon(\lambda):=\lambda^2\int_{\T^3}
\frac{d^3p}{e(p)} +\lambda^{4-\epsilon}\,,\end{equation} then
$E^*>C\lambda^{4-\epsilon}$ for an arbitrary small $\epsilon$ and
sufficiently small values of $\lambda$. We therefore can define a
(renormalized) free Green function
\begin{equation}\label{eq:R_r}R_r(x,y)=(-\frac{1}{2}\Delta+E-\sigma(E)+i0)^{-1}(x,y)\end{equation} for
every $E$ in the above energy range. Let us also denote by
\begin{equation}\label{eq:R}R(x,y)=(H^\lambda_\omega+E+i0)^{-1}(x,y)\,,\end{equation} which is well defined
a.s., \cite{ASFH}.

The hallmark of localization is rapid decay of $G(x,y)$ at energies
in the spectrum of $H_\omega$, for the typical configuration
$\omega$. Rapid decay of the Green function is related to the
non-spreading of wave packets supported in the corresponding energy
regimes and various other manifestations of localization whose
physical implications have been extensively studied in regards to
the conductive properties of metals and in particular to the quantum
Hall effect. Our main result (Theorems \ref{thm:main} and \ref{cor}
below) establishes this behavior of the Green function at the band
edges of the spectrum, by comparing it with the asymptotics of the
free Green function $R_r(x,y)$. The behavior of the latter for
$|x-y|\gg1$ is known \cite{KI}:
\[R_r(x,y)\sim\frac{1}{2\pi |x-y|}\,e^{-\sqrt{2E^*}|x-y|}\,.\]
The implication is that the correlation length for the free Green
function is $(E^*)^{-1/2}$. Moreover, for the energy range
(\ref{eq:range}) we have
\[(|x-y|+1)^{-1}\gg \lambda((|x-y|+1)^{-1/2}\] whenever
\[|x-y|<(E^*)^{-1/2}\,,\] hence
\[|R_r(x,y)|^s\gg \frac{\lambda^s}{(|x-y|+1)^{s/2}}\,,\quad {\rm for }
\ |x-y|<(E^*)^{-1/2}\,,\quad E\ge E_\epsilon(\lambda)\,.\]  With
this estimate in mind, we present
\begin{thm}[Local fractional moment bound]\label{thm:main}
For $H_\omega^\lambda$ as above, for any $s<1/2$ and $\epsilon>0$,
there exists $\lambda_0(\epsilon)$  such that for all
$\lambda<\lambda_0(\epsilon)$ and $E\ge E_\epsilon(\lambda)$, one
has a bound
\begin{equation}\label{eq:main}
 \E\ |R(x,y)-R_r(x,y)|^s \
\le \ C_1(s)\, \frac{\lambda^{s}}{(|x-y|+1)^{s/2}}\,,
\end{equation}
with $C_1(s)<\infty$, which holds for any pair $\{(x,y)\in
\Z^3\times\Z^3:\ |x-y|<(E^*)^{-1/2}\}$.
\end{thm}
The estimate (\ref{eq:main}) can be interpreted as follows: The
typical correlation length $\xi_E$ within this energy range cannot
be smaller than $O((E^*)^{-1/2})$, in particular it grows as $E^*$
approaches zero\footnote{Unfortunately, in this work we are only
able to descend to the values $E^*=O(\lambda^{4-\epsilon})$, so we
cannot claim that the correlation length indeed diverges.}.
 The next proposition shows that $\xi_E$ cannot be of much greater scale
either.
\begin{thm}[Global fractional moment bound]\label{cor}
For $H_\omega^\lambda$ as above, there exists $\lambda_0(\epsilon)$,
so that for all $\lambda<\lambda_0(\epsilon)$, $s<1/4$, and $E\ge
E_\epsilon(\lambda)$, one has a bound
\begin{equation}\label{eq:main1}
\E\ |R(x,y)|^s \ \le \ \frac{C_2(s)}{\lambda^s}\,
e^{C_3\sqrt{E^*}\ln({E^*})|x-y|}
\end{equation}
for all $x,y\in\Z^3$.
\end{thm}
Let us list several  known implications of the global fractional
moment bound:
\begin{itemize}
\item[i.]  {\it Spectral localization
(\cite{AM}):} The spectrum of $H_{\omega}$ within the interval
(\ref{eq:range}) is almost-surely of the pure-point type, and the
corresponding eigenfunctions are exponentially localized.

\item[ii.] {\it Dynamical localization (\cite{A}):}
Wave packets with energies in the specified range do not spread (and
in particular the {\em SULE} condition of \cite{SULE} is met):
\begin{equation} \E\left( \sup_{t\in \R} \left|\left(
P_{\{H^\lambda_\omega<-E_\epsilon\}}\,e^{-i tH}
\right)(x,y)\right|\right) \ \le \widetilde A \e^{-\tilde \mu |x-y|}
\; , \label{eq:dyn}
\end{equation}
where $P_{H<a}$ stands for the spectral projection of $H$ on the
energies below $a$.
\item[iii.] {\it Absence of level repulsion (\cite{Min}).}
Minami  has shown that \eqref{eq:main1} implies that in the range
(\ref{eq:range}) the energy gaps have Poisson-type statistics.

\end{itemize}
For energies $E$ slightly above $-E_0$ it is expected on the
physical grounds (see also a discussion below) that
$H^\lambda_\omega$ should almost surely have absolutely continuous
spectrum in $3D$. Presumably, the correlation length truly diverges
when one starts to approach the mobility edge and it will be
extremely interesting to cover the missing case  of
$E\in[-E_\epsilon(\lambda), -E_0]$.

This result can also be established by studying the density of
states. Once the DOS is shown to be small below $E^*$, the
localization is a rather straightforward consequence of known
methods (as say in \cite{Klopp}). However, if one wants to study the
behavior of the model at the closer vicinity of $E^*$, this extra
step can be an obstacle, as DOS increases.

As was pointed to us by the referee, the numerical results seem to
qualitatively agree with the suggestion that the mobility edge is
near $ - C\lambda^2$ (e.g. \cite{SZE}). If one uses Cauchy variables
instead of the box distribution, it is possible to calculate certain
quantities explicitly for such variables: $\E(f(v)) = f(i)$ for a
function $f$ having a bounded analytic continuation to the upper
half plane. In particular, DOS is computable, and it is not small
for negative energies (namely $O(\sqrt\lambda)$). The lower mobility
edge in $d=3$ appears to be positive on the basis of numerical
studies (e.g. \cite{KK}, Figure 5), which suggests that the
existence of high moments of the distribution of the potential plays
a crucial role in the analysis.

\subsection{Major steps in the proof}

In the unpublished notes, T. Spencer \cite{Spencer} proposed to
prove the localization near the band edge using  the multiscale
analysis, with the initial volume estimates coming from the fact
that the the  density of states in the Lifshitz tail regime is
small.  To control the density of states he suggested to truncate
the resolvent expansion at some optimal point. The corresponding
Feynman graphs become superficially convergent after the suitable
renormalization. In this paper, we complete the proof of the result
announced in \cite{Spencer}, combining Spencer's perturbative
approach with the Aizenman-Molchanov fractional moment method and
developing the detailed estimates on the error terms in the
renormalized expansion.

The following representation for a Green function $R(x,y)$ will be
useful:
\begin{lemma}\label{lemma:rep}
 For any integer $N$ and energies $E$ that
satisfy (\ref{eq:range}) we have the decomposition
\begin{equation}\label{eq:mdec}
R(x,y)\ = \ \sum_{n=0}^{N-1}A_n(x,y) \ + \ \sum_{z\in Z^3}\tilde
A_N(x,z)R(z,y)\,,
\end{equation}
with $A_0(x,y)=R_r(x,y)$, and where the (real valued) kernels
$A_n,\,\tilde A_N$ satisfy bounds
\begin{eqnarray}
\E\, (A_n(x,y))^2\ \le \ (4n)!\, E^*\,\left(C(E^*)\
\frac{\lambda^{2}}{\sqrt{E^*}}\right)^n\,e^{-\sqrt\frac{{E^*}}{3}\,|x-y|}\,,\
 \ n>1 \,;\label{eq:l}
\\
\E\, |\tilde A_N(x,y)|\ \le \ \sqrt{(4N)!}\,\left(C(E^*)\
\frac{\lambda^{2}}{\sqrt{E^*}}\right)^{N/2}\,e^{-\sqrt\frac{{E^*}}{12}|x-y|}\,,\
 \ N>1 \,;\label{eq:lt}
\end{eqnarray}  where $C(E^*)=K\ln^{9}E^*$ for some generic constant
$K$.

The zero order contribution $A_0$ satisfies
\begin{equation}\label{eq:stand}
0 \ <\ A_0(x,y)=\int_{\T^3}e^{i(x-y)p}\frac{d^3p}{e(p)+E^*}\ \le \
\frac{K}{(|x-y|+1)}
\end{equation}
for all $x,y\in\Z^3$, and behaves asymptotically as
\begin{equation}\label{eq:asymp}
A_0(x,y) \ = \
\left(1+O\left(\sqrt{E^*}\right)+O\left(|x-y|^{-1}\right)\right)\times\frac{e^{-\sqrt{2E^*}|x-y|}}{2\pi
(|x-y|+1)} \,.
\end{equation}
Lastly, we have \begin{equation}\label{eq:a1}
 \E\, (A_1(x,y))^2\
\le \ \frac{ K\, \lambda^2}{|x-y|+1}\,e^{-2\sqrt{2E^*}|x-y|} \,.
\end{equation}
\end{lemma}
One then looks for the optimal value $N$ to stop the expansion -
note that the increasing  factor of $(4N)!$ in $A_N(x,y)$ competes
with the decreasing factor $(\lambda^4 E^*)^{N/2}$.

The choice $\E^*>\lambda^{4-\epsilon}$ has the effect that
\begin{equation}\label{eq:refe}C(E^*)\ \frac{\lambda^{2}}{\sqrt{E^*}} \ \le \
\lambda^{B\epsilon}\,,\quad 0 < B < 1\,,\end{equation} which
suffices to control \eqref{eq:l} -- \eqref{eq:lt}. It turns out that
the appropriate choice for N should satisfy
\[(4N)!\,\left(\frac{C(E^*)\lambda^2} {\sqrt{E^*}}\right)^{N}\approx
e^{-N}\] (see the next section for details). In terms of the
$\lambda$ - dependence, it corresponds to
$N\sim\lambda^{-b\epsilon}$ for $b < B$. Note that the square root
in the denominator of \eqref{eq:refe} is absolutely crucial for the
strategy. To this end, let us mention that the representation
\eqref{eq:mdec} is a resolvent type expansion (cf. Lemma
\ref{lem:de} below). If one applies the rough norm bound on a each
factor of the resolvent there, the denominator in \eqref{eq:refe}
will contain $E^*$ rather than its square root. The improvement is
achieved using the Feynman diagramatic technique (Section
\ref{section:lemma}).

Let us denote by $H_\omega^{\Lambda,\lambda}$ the natural
restriction of $H_\omega^{\lambda}$ to  $\Lambda\subseteq\Z^3$ and
let $R^\Lambda=(H_\omega^{\Lambda,\lambda}+E+i0)^{-1}$ be the
corresponding resolvent.

Theorems \ref{thm:main} and \ref{cor} follow from the result above
and from Aizenman--Molchanov a-priori bound on the fractional moment
of the Green function as it appears in Lemma 2.1 of \cite{ASFH},
which states that
\begin{equation}\label{eq:apriori}
\E\, |R^\Lambda(x,y)|^s\ < \ C_s
\end{equation}
for any $0<s<1$, uniformly in $x,y\in \Lambda$ and $\lambda$.
Moreover, the bound above holds uniformly for an arbitrary set
$\Lambda$.

The rest of the paper is organized as follows: To make the
presentation less obscure, we postpone the rather lengthy proof of
 the main technical Lemma \ref{lemma:rep} until Section
\ref{section:lemma} and establish first Theorems \ref{thm:main} and
\ref{cor}. In Section \ref{section:renorm} we perform a self energy
renormalization required to get rid of the so called tadpole
contributions. A technical statement regarding the properties of the
self energy term $\sigma(E)$ is proven in Appendix
\ref{sec:appendI}.
\newpage

\section{Proofs of Theorem \ref{thm:main} and \ref{cor}}
\subsection{Theorem \ref{thm:main}}
Lemma \ref{lemma:rep}, Aizenman--Molchanov a-priori bound
(\ref{eq:apriori}), and H\"older inequality imply that for any
$0<s<1/2$
\begin{multline}
\E\, |R(x,y)-R_r(x,y)|^s \\ \le \ \sum_{l=1}^{N-1}\E\,A_l^{s}(x,y) \
+ \ \sum_{z\in Z^3}\left(\E\,|\tilde
A_N(x,z)|^{2s}\right)^{1/2}\left(\E\,|R(z,y)|^{2s}\right)^{1/2}
\\ \le \ \sum_{l=1}^{N-1}\left(\E\,A_l^2(x,y)\right)^{s/2} \ + \
\sum_{z\in Z^3}\left(\E\,|\tilde
A_N(x,z)|\right)^{s}\left(\E\,|R(z,y)|^{2s}\right)^{1/2}
\\ \le \ \frac{K^{s/2}\lambda^s}{|x-y|^{s/2}+1}\,e^{-s\sqrt{2E^*}\,|x-y|} \\ + \
\sum_{l=2}^{N-1}\,((4l)!)^{s/2}\,(E^*)^{s/2}\left(\frac{C(E^*)\lambda^2}
{\sqrt{E^*}}\right)^{sl/2}\,e^{-s\sqrt{\frac{E^*}{12}}\,|x-y|} \\
+ \ C(s) ((4N)!)^{s/2}\,\left(\frac{C(E^*)\lambda^2}
{\sqrt{E^*}}\right)^{sN/2}\,\sum_{z\in
Z^3}\,e^{-s\sqrt{\frac{E^*}{12}}|x-z|}
\\ \le \
 \frac{K^{s/2}\lambda^s}{|x-y|^{s/2}+1}\,e^{-s\sqrt{2E^*}\,|x-y|} \\ + \
(C(E^*)\lambda^2)^{s}\,e^{-s\sqrt{\frac{E^*}{12}}\,|x-y|}\,
\sum_{l=2}^{N-1}\,((4l)!)^{s/2}\,\left(\frac{C(E^*)\lambda^2}
{\sqrt{E^*}}\right)^{s(l-2)/2} \\ + \ \frac{\tilde
C(s)}{(E^*)^{3/2}}((4N)!)^{s/2}\,\left(\frac{C(E^*)\lambda^2}
{\sqrt{E^*}}\right)^{sN/2}\ \,.
\end{multline}
Choosing \[(4N)^4=\frac{\sqrt{E^*}}{C(E^*)\lambda^2} \] one obtains,
using the  Stirling's approximation, that the summation over the
index $l$ is bounded by some $s$ - dependent constant. On the other
hand, for such $N$ we have $(4N)!\,\left(\frac{C(E^*)\lambda^2}
{\sqrt{E^*}}\right)^{N}< e^{-N}$. Hence, for such a value of $N$ we
have
\begin{multline}\label{eq:infbnd}
\E\, |R(x,y)-R_r(x,y)|^s  \ < \
C_s\,\left(\frac{\lambda^s\,e^{-s\sqrt{2E^*}\,|x-y|}}{|x-y|^{s/2}+1}\right.
\\ \left.\ + \ (C(E^*)\lambda^2)^{s}\,e^{-s\sqrt{\frac{E^*}{12}}\,|x-y|} \ +
\
(E^*)^{-3/2}\,\exp{\left(-\frac{s\,\sqrt[4]{\frac{\sqrt{E^*}}{C(E^*)\,\lambda^2}}}{8}
\right)} \right)
\end{multline}
with some generic constant $C_s$. We infer that for any $\epsilon>0$
and sufficiently small $\lambda_0(\epsilon)$ we have for any
$\lambda<\lambda_0(\epsilon)$ and $E^*>\lambda^{4-\epsilon}$
\begin{equation}\label{eq:loc}
\E\, |R(x,y)-R_r(x,y)|^s \ \le \ C_s\
\frac{\lambda^s}{|x-y|^{s/2}+1}\,,\quad |x-y|<(E^*)^{-1/2}\,,
\end{equation}
(that is we proved Theorem \ref{thm:main}) and
\begin{equation}\label{eq:glob}
\E\, |R(x,y)|^s \ \le \ C'_s\
\left(\lambda^{s}\,e^{-s\,\frac{\sqrt{E^*}}{4}\,|x-y|}  \ + \
e^{-s\, \lambda^{-\epsilon/9}}\right)\,,
\end{equation}
for any $x,y\in\Z^3$, where we have used the bound (\ref{eq:asymp})
for $R_r(x,y)$. The latter estimate will be used in the proof of
Theorem \ref{cor}.


\subsection{Theorem \ref{cor}} Using notation introduced after
(\ref{eq:apriori}) we define the decoupled Hamiltonian $H_\Lambda$
to be:
\[H_\Lambda=H_\omega^{\Lambda,\lambda}\oplus H_\omega^{\Lambda^c,\lambda}\,,\]
where $\Lambda$ is a cubic box of the linear size $2L$ centered
around the origin, and $\Lambda^c:=\Z^3\setminus\Lambda$. Let
$\partial\Lambda$ be  a boundary  $\Lambda$. We will denote
$R^\Lambda:=(H_\Lambda+E^{*})^{-1}$. For any $s<1/2$ and
$n\in\partial\Lambda$ we have
\begin{multline}\label{eq:ini}
\E\,|R^\Lambda(n,0)|^s \ \le \ \E\,|R(n,0)|^s \ + \
\E\,|R^\Lambda(n,0)-R(n,0)|^s\\ = \ \E\,|R(n,0)|^s \ + \
\E\,|\left(R^\Lambda\,
(H_\Lambda-H_\omega^\lambda)\,R\right)(n,0)|^s
\\ \le
\ \E\,|R(n,0)|^s \\ + \
\sum_{\dist(k,\partial\Lambda)\le1}\Big\{\E\,| \left(R^\Lambda\,
(H_\Lambda-H_\omega^\lambda)\, \right)(n,k)|^{2s}\Big\}^{1/2}
\,\Big\{\E\,|R(k,0)|^{2s}\Big\}^{1/2}\,,
\end{multline}
where we used locality of  $(H_\Lambda-H_\omega^\lambda)$ - its
non-zero matrix elements lies essentially on the boundary of
$\Lambda$. Similar considerations lead to the estimate
\begin{equation}\label{eq:diff}
\E\,|\left(R^\Lambda(H_\Lambda-H_\omega^\lambda)\right)(n,k)|^{2s} \
\le \ C\,
\sup_{m:\dist(m,\partial\Lambda)\le1}\E\,|R^\Lambda(n,m)|^{2s}\,.
\end{equation}
Using bounds (\ref{eq:diff}) and (\ref{eq:apriori}) as well as the
 H\"older inequality, we obtain from (\ref{eq:ini})
\begin{equation}
\E\,|R^\Lambda(n,0)|^s \ \le \ C_s
\sum_{k\in\partial\Lambda}\Big\{\E\,|R(k,0)|^{2s}\Big\}^{1/2}\ \,.
\end{equation}
Plugging the bound (\ref{eq:glob}) into the latter equation, with
$s<1/4$, we establish
\begin{equation}\label{eq:pre}
\E\,|R^\Lambda(n,0)|^s \ \le \
C_s\,L^2\left(\lambda^s\,e^{-sL\sqrt{\frac{E^*}{12}}} \ + \
e^{-s\lambda^{-\epsilon/9}}\right)
\end{equation}
Now we are in a position to use the fractional moment criterion
\cite{ASFH}, Theorem 1.2, which states that if
\begin{equation}\label{eq:fmc}
B_s L^4
\lambda^{-2s}\,\sum_{n\in\partial\Lambda}\E\,|R^\Lambda(n,0)|^s \ <
\ b \,,
\end{equation}
for a certain constant $B_s$ and $b<1$, then we have a bound
\begin{equation}\label{eq:main2}
\E\ |R(x,y)|^s \ \le \ \frac{B_s}{b^2\lambda^s}\, e^{\frac{\ln
b}{L}|x-y|}\,.
\end{equation}
It is easy to see  from (\ref{eq:pre}) that  (\ref{eq:fmc}) is
satisfied, provided that $E^*>\lambda^{4-\epsilon}$, and $\lambda$
is small enough\footnote{Since the bottom of the spectrum of
$H_\omega$ is almost surely located at $\lambda a$, we can assume
without loss of generality that $E^*\le|a|\lambda$.}, with
\[L=O\left(\frac{\ln(E^*)^{-1}}{s\sqrt{E^*}}\right)\,,\] hence the result.


\section{Renormalization of tadpole's
contribution}\label{section:renorm}
\subsection{Expectation of the product of the random
potentials}\label{sub:prod} In what follows, the expectation of the
product of random variable $V_\omega$, namely \[\E
\left[\prod_{j=1}^n V_\omega(x_j)\right]\] will play an important
role. These products naturally  arise when one starts to expand the
operator $R:=(H^\lambda_\omega + E+i0)^{-1}$ in the resolvent series
\[R=\sum_{i=0}^n(-\lambda R_0 V_\omega)^iR_0\ + \
(-\lambda R_0 V_\omega)^{n+1}R\] about the (unperturbed) operator
$R_0:=(H_0 + E)^{-1}$. Indeed,
\begin{multline}\label{eq:gate}
(-\lambda R_0 V_\omega)^nR_0(x_0,x_{n+1})  =  (-\lambda )^n
\sum_{x_j\in\Z^3;\,j=1,..n}
\prod_{j=1}^nV_\omega(x_j)\prod_{i=0}^nR_0(x_i,x_{i+1})\,.
\end{multline}
Let $\Upsilon_{N,N'}$ be the set $\{1,...,N,N+2,...N+N'+1\}$, while
$\Pi_{N,N'}$ will denote the set of partitions of $\Upsilon_{N,N'}$
into disjoint subsets $S_j$ of size $|S_j|\in 2\N$. Two
partitions $\pi = \{S_j\}_{j=1}^m$, $\pi' = \{S'_j\}_{j=1}^m$  are
equivalent, $\pi=\pi'$, if they coincide up to the permutation. For
$S\subset \Upsilon_{N,N'}$, let
\begin{equation}
\delta(x_{S})=\sum_{y\in\Z^3}\prod_{j\in S}\delta_{|x_j-y|}\,,
\end{equation}
where $\delta_{x}$, $x\in \Z$ is Kronecker delta function, and $x_S$
denotes the collection of $\{x_i\,,\ i\in S\}$.
 One has an
identity (see e.g. \cite{Chen} Section 3.1 for details)
\begin{equation}\label{eq:tree_graph}
\E \left[\prod_{j\in \Upsilon_{N,N}}
V_\omega(x_j)\right]=\sum_{m=1}^N\sum_{\pi=\{S_j\}_{j=1}^m}\,
\prod_{j=1}^m c_{|S_j|}\delta(x_{S_j})\,,
\end{equation}
with coefficients $c_{2l}\le (cl)^{2l+1}$ proviso (A1-A2), and
$c_2=\E\, (V_\omega^2(x))=1$. The set $S_j$ in the partitions
$\pi\in \Pi_{N,N'}$ can be of the special type: If
\begin{equation}\label{eq:gatedef}
 S_j=\{i,i+1\}
\end{equation}
we will refer to it as a {\em tadpole}, or a {\em gate} set. The non
zero order contributions in the resolvent expansion, associated with
the tadpole--free terms, are sufficiently small in the energy range
(\ref{eq:range}), as opposed to the contributions containing the
gates.
\subsection{Self energy renormalization}
The purpose of renormalization  is then to include the tadpole
contributions  into the propagator itself. In our case this can be
established by subtracting from the unperturbed operator
$-\frac{1}{2}\Delta$ the  self-energy term $\sigma(E)$, described in
Section \ref{sec:intro}. We  decompose $H_\omega^\lambda$ as
\[H_\omega^\lambda=H_r+\tilde V\,,\quad
H_r:=-\frac{1}{2}\Delta-\sigma(E)\,,\quad \tilde V:=\lambda V_\omega
+ \sigma(E)\,.\] The corresponding resolvent expansion for $R$
defined in (\ref{eq:R}) takes the form
\begin{equation}\label{eq:expa}R=\sum_{i=0}^n(-R_r\tilde V)^iR_r\ + \
(-R_r\tilde V)^{n+1}R\,,\end{equation} with $R_r$ as in
(\ref{eq:R_r}). Note that for any $x\in\Z^3$
\begin{equation}\label{eq:bulle}
\sigma(E)=\lambda^2\, R_r(x,x)\,.
\end{equation}
In place of (\ref{eq:gate}) we get
\begin{equation}\label{eq:gate1}(-\lambda R_r \tilde V)^nR_r(x_0,x_{n+1}) \ = \
\sum_{x_j\in\Z^3;\,j=1,..n} \prod_{j=1}^n(-\lambda
V_\omega(x_j)-\sigma(E))\prod_{i=0}^nR_r(x_i,x_{i+1})\end{equation}
If we open the brackets in (\ref{eq:gate1}), we obtain
\[
\sum_{\theta,\,x_j\in\Z^3;\,j=1,..n}R_r(x_0,x_1)\theta(x_1)R_r(x_1,x_2)\theta(x_2)...R_r(x_{n-1},x_n)
\theta(x_n)R_r(x_n,x_{n+1})
\]
where $\theta(x)$ is either $-\lambda V_\omega(x)$, or $-\sigma(E)$
(whenever $\theta(x)=-\sigma(E)$ we will refer to it as a {\em
bullet}). Since $\sigma(E)=O(\lambda^2)$ for all permissible values
of $E$, see Appendix \ref{sec:appendI}, one can unambiguously define
the {\em order} $l$ (in powers of $\lambda$) of the particular
contribution
\[R_r(x_0,x_1)\theta(x_1)R_r(x_1,x_2)\theta(x_2)...R_r(x_{n-1},x_n)
\theta(x_n)R_\sharp(x_n,x_{n+1})\,,\] (with $R_\sharp$ being either
$R_r$ or $R$) according to the following rule: Each factor of
$\sigma(E)$ counts as $2$, while appearance of the random potential
counts as $1$, and we add up all the exponents to get the order of the
term. For instance, the order of the expression
\[R_r(x_0,x_1)\sigma(E)R_r(x_1,x_2)\lambda V_\omega(x_2)R_r(x_2,x_3)
\sigma(E)R(x_3,x_4)\] is $5$.

To handle the renormalization of tadpole contributions properly, we
decide at which value of $n$ to halt the expansion in
(\ref{eq:expa}) individually for each contribution according to the
following rule (to which we will refer as a stopping rule): If the
order of the kernel
\begin{equation}
R_r(x_0,x_1)\theta(x_1)R_r(x_1,x_2)\theta(x_2)...R_r(x_{n-1},x_n)
\theta(x_n)R(x_n,x_{n+1})
\end{equation}
reaches or exceeds a value $N$ to be determined later on, we stop
expanding this particular term. To illustrate this procedure we
write down the expansion obtained in a case of $N=2$:
\begin{multline}R\ = \ R_r \ - \
R_r\sigma(E)R \
- \ \left\{\lambda R_r V_\omega R \right\} \ = \\
 R_r \ - \ R_r\sigma(E)R\ - \ \lambda
R_rV_\omega R_r \\ + \ \lambda R_r V_\omega R_r \sigma(E)R \
  + \ \lambda^2 R_r
V_\omega R_r V_\omega R  \,,
\end{multline} where the term in the curled brackets is the one
we expanded according to the stopping rule. Note that the
penultimate term is of order $3$. It is not difficult to see that
for a general $N$ we get:
\begin{lemma}\label{lem:de}
For any integer $N$ we have a decomposition  (used in Lemma
\ref{lemma:rep})
\begin{multline}\label{eq:gra}
R(x,y)\ = \ \sum_{z\in Z^3}\Big(\sum_{l=0}^{N-1}
A'_l(x,z)R_r(z,y)+A'_N(x,z)R(z,y)+B_N(x,z)R(z,y)\Big)
\\  = \ \sum_{l=0}^{N-1}A_l(x,y) \ + \ \sum_{z\in Z^3}\tilde
A_N(x,z)R(z,y)\,,
\end{multline}
where $A'_0(x,z)=\delta_{|x-z|}$, $A'_l(x,z)$ is a summation over
all possible terms of the type
\begin{equation}\label{eq:ord}
\sum_{\theta,\,x_j\in\Z^3;\,j=1,..n}R_r(x,x_1)
\theta(x_1)R_r(x_1,x_2)\theta(x_2)...R_r(x_{n},z) \theta(z)
\end{equation}
which are of the order $l>0$, while
\begin{equation}\label{eq:ord1}B_N(x,z)=-\sigma(E)\sum_{w\in Z^3}
A'_{N-1}(x,w)R_r(w,z)\,.
\end{equation}
The quantities $A_l$ and $\tilde A_N$ are defined as
\[A_l(x,y)=\sum_{z\in Z^3}A'_l(x,z)R_r(z,y)\,,\quad \tilde
A_N(x,y)=A'_N(x,y)+B_N(x,y)\,.\]
\end{lemma}
\begin{proof}
Note first that it is follows from (\ref{eq:ord}, \ref{eq:ord1})
that
\begin{equation}\label{eq:ord2}
A_{N+1}(x,y)= B_N(x,y)-\lambda\sum_{z,w\in Z^3}A'_N(x,z)
R_r(z,w)V_\omega(w)R(w,y)\,.\end{equation} We now prove
(\ref{eq:gra}) by induction: The base of induction, $N=0,1$, gives
equalities \begin{multline}R(x,y)=R(x,y),\\
R(z,y)=R_r(z,y)-\sum_{w\in Z^3}\Big(\lambda
R_r(z,w)V_\omega(w)R(w,y)+\sigma
R_r(z,w)R(w,y)\Big)\,.\nonumber\end{multline} Suppose that
(\ref{eq:gra}) holds for $N$, then
\begin{multline}
R(x,y)\ = \ \sum_{z\in Z^3}\Big(\sum_{l=0}^{N-1}
A'_l(x,z)R_r(z,y)+A'_N(x,z)R(z,y)+B_N(x,z)R(z,y)\Big)
\\  = \ \sum_{z\in Z^3}\Big(\sum_{l=0}^{N-1}
A'_l(x,z)R_r(z,y)+A'_N(x,z)R_r(z,y)+B_N(x,z)R(z,y)\Big)
\\  - \ \sum_{z,w\in Z^3}A'_N(x,z)\Big(
\lambda R_r(z,w)V_\omega(w)R(w,y)+\sigma R_r(z,w)R(w,y)\Big)\\  = \
\sum_{z\in
Z^3}\Big(\sum_{l=0}^{N}A'_l(x,z)R_r(z,y)+A'_{N+1}(x,z)R(z,y)+B_{N+1}(x,z)R(z,y)\Big)\,,
\end{multline}
where in the last line we used (\ref{eq:ord1}, \ref{eq:ord2}).
\end{proof}
Such a stopping procedure guarantees cancelation of tadpoles and
bullets in the following sense:
\begin{lemma}
We have
\begin{multline}\label{eq:gaf1}
\E\, A^2_l(x,y)\ = \ \lambda^{2l}\, \sum_{m=1}^{l}  \,
\sum_{\{S_j\}_{j=1}^m}^{\ \ \ \ \ '}\,
\sum_{\left(\Z^3\right)^{2l}}\,R_r(x,x_1)R_r(x,x_{l+2})\,
\prod_{i\in\Upsilon_{l,l}}R_r(x_i,x_{i+1})\\ \times \ \prod_{j=1}^m
c_{|S_j|}\delta(x_{S_j})\,
 \,,
\end{multline}
with $\sum_{\left(Z^3\right)^{2l}}$ standing for summation over
$\Z^3$ of all variables $x_j$ with $j\in\Upsilon_{l,l}$,
$x_{l+1}=x_{2l+2}=y$, and where  $\sum'$  denotes summation over all
possible partitions of $\Upsilon_{l,l}$ which do not contain gates.
\par Equivalently, in the momentum representation
\begin{multline}\label{eq:gaf}
\E\, A_l^2(x,y)\ = \ \lambda^{2l}\,\int  \, e^{i\alpha}\
\frac{dp_{l+1}}{e(p_{l+1})+E^*}\,\frac{dp_{2l+2}}{e(p_{2l+2})+E^*}\
\\ \times \prod_{t\in\Upsilon_{l,l}} \,\frac{dp_t}{e(p_t)+E^*}\,
\sum_{m=1}^l\sum_{\pi=\{S_j\}_{j=1}^m}^{\ \ \ \ \ '} \prod_{j=1}^m
c_{|S_j|}\,\delta(\sum_{i\in S_j}p_i-p_{i+1}) \,,
\end{multline}
where  \[ \alpha:=2\pi\{-(p_1+p_{l+2})\cdot x
+(p_{l+1}+p_{2l+2})\cdot y\}\,.\]
\end{lemma}
\remark In order to obtain (\ref{eq:gaf}) from (\ref{eq:gaf1}) one
uses
\[R_r(z,w)\ = \ \int_{\T^3}e^{i2\pi(z-w)p}\frac{d^3p}{e(p)+E-\sigma(E)}
\ =  \ \int_{\T^3}e^{i2\pi(z-w)p}\frac{d^3p}{e(p)+E^*}\,.\]

\begin{proof}
Let us introduce some extra notation.  Let $\Pi_{l,l;k}$ denote the
set of partitions of $\Upsilon_{l,l}$ into disjoint subsets $S_j$
such that each of these partitions contains exactly $k$ tadpoles
(that is there are $k$ subsets $S_j$ of the form $S_j=\{i,i+1\}$).
For every partition $\Pi_{l,l;k}\ni\pi_k= \{S_j\}_{j=1}^m$ we will
denote by $\Upsilon(\pi_k)$ a subset of $\Upsilon_{l,l}$ which
consists of the indices that differ from the set
$\Upsilon^c(\pi_k):=\Upsilon_{l,l}\backslash\Upsilon(\pi_k)$ of the
gate's indices associated with $\pi_k$ (so ${\rm
card}(\Upsilon(\pi_k))=2l-2k$). Let
$\hat\pi_k=\{S_j\}_{\{j:S_j\subseteq\Upsilon(\pi_k)\}}$. Let
$\nu(v)$ be the connected segment of the set of the indices
$\Upsilon^c(\pi_k)$ which contains index $v$, that is
\[v\in\nu(v)\subseteq \Upsilon^c(\pi_k)\] and
$\nu(v)=\{i,i+1,...,h-1,h\}$, with $i-1,h+1\in\Upsilon(\pi_k)$. Let
$d(v)$ denote the position of index $v$ with respect to $\nu(v)$,
for example, if $\nu(4)=\{3,4,5,6,7,8\}$, then $d(4)=2$. We define
$\hat{\prod}_{v\in\Upsilon^c(\pi_k)}$ to be a product over such
$v\in\Upsilon^c(\pi_k)$ that $d(v)\mod 2 =1$. For instance, if
$\Upsilon^c(\pi_k) = \{1,2,4,5,6,7\}$, the product will run over the
variables $1,4,6$.

We can  now express $A_l(x,y)^2$ as
\begin{multline}\label{eq:re}
A_l^2(x,y)\ = \ \sum_{k=0}^l\,\lambda^{2l-2k}(-\sigma(E))^k\,
\sum_{\pi_k\in\Pi_{l,l;k}}\, \sum_{\left(\Z^3\right)^{2l}}\,R_r(x,x_1)R_r(x,x_{l+2})\,\\
\times\ \prod_{i\in\Upsilon(\pi_k)}V_\omega(x_{i})R_r(x_i,x_{i+1}) \
\hat{\prod_{j\in\Upsilon^c(\pi_k)}}R_r(x_{j+1},x_{j+2})\delta_{|x_{j+1}
- x_{j}|}
 \,.
\end{multline}
Note that here the index $k$ corresponds to the number of bullets in
the corresponding contribution, and is not related to the number of
the tadpoles (which will show up as index $k'$ below once we
undertake the expectation over disorder).

On the other hand, since $c_2=1$, one obtains an identity
\begin{multline}\label{eq:gates}
\sum_{m=1}^N\sum_{\pi=\{S_j\}_{j=1}^m}\, \prod_{j=1}^m
c_{|S_j|}\delta(x_{S_j})\\ = \sum_{k'=0}^N \sum_{m=1}^{N-k'} \,
\sum_{\hat\pi_{k'}=\{S_j\}_{j=1}^m}\prod_{j=1}^m
c_{|S_j|}\delta(x_{S_j})\ \hat{\prod_{v\in\Upsilon^c(\pi_{k'})}}\,
\delta_{|x_{v+1}-x_v|}\,.
\end{multline}
At this point, let us first compute  the expectation of $k=0$ part
of the summation in (\ref{eq:re}):
\begin{multline}\label{eq:expec}
\lambda^{2l}\, \sum_{\pi_0\in\Pi_{l,l;0}}\,
\sum_{\left(\Z^3\right)^{2l}}\, R_r(x,x_1)R_r(x,x_{l+2})\, \E
\left[\prod_{i\in\Upsilon(\pi_0)}V_\omega(x_{i})R_r(x_i,x_{i+1})\right]
\\ = \  \lambda^{2l} \sum_{{k'}=0}^l
\sum_{m=1}^{l-{k'}} \, \sum_{\hat\pi_{k'}=\{S_j\}_{j=1}^m}\,
\sum_{\left(\Z^3\right)^{2l}}\,
R_r(x,x_1)R_r(x,x_{l+2})\,\prod_{j=1}^m
c_{|S_j|}\delta(x_{S_j})\\
\times \ \prod_{i\in\Upsilon(l,l)}R_r(x_i,x_{i+1})\
\hat{\prod_{v\in\Upsilon^c(\pi_{k'})}}\, \delta_{|x_{v+1}-x_v|}\\ =
\
  \sum_{{k'}=0}^l\lambda^{2l-2{k'}}(\sigma(E))^{k'}
\sum_{m=1}^{l-{k'}} \, \sum_{\hat\pi_{k'}=\{S_j\}_{j=1}^m}\,
\sum_{\left(\Z^3\right)^{2l}}\,R_r(x,x_1)R_r(x,x_{l+2})\,\\
\times\ \prod_{j=1}^m c_{|S_j|}\delta(x_{S_j})\,
\prod_{i\in\Upsilon(\pi_{k'})}R_r(x_i,x_{i+1}) \
\hat{\prod_{j\in\Upsilon^c(\pi_{k'})}}R_r(x_{j+1},x_{j+2})\delta_{|x_{j+1}
- x_{j}|} \,,
\end{multline}
where we have used $\lambda^2 R_r(z,z)=\sigma(E)$ for all
$z\in\Z^3$.

More generally, we have an equality
\begin{multline}
\lambda^{2l-2k}(-\sigma(E))^k\,
\sum_{\pi_k\in\Pi_{l,l;k}}\, \sum_{\left(\Z^3\right)^{2l}}\,R_r(x,x_1)R_r(x,x_{l+2})\,\\
\times\ \E
\left[\prod_{i\in\Upsilon(\pi_k)}V_\omega(x_{i})R_r(x_i,x_{i+1})\right]
\
\hat{\prod_{j\in\Upsilon^c(\pi_k)}}R_r(x_{j+1},x_{j+2})\delta_{|x_{j+1}
- x_{j}|}
\\ =   \sum_{k'=0}^{l-k}\, \Big(_{\ \ \,k'\,\ }^{
\,k'+k\,}\Big)\,
\lambda^{2l-2k-2k'}\,(-\sigma(E))^k\,(\sigma(E))^{k'}
\\ \times\ \sum_{m=1}^{l-k-k'} \,
\sum_{\hat\pi_{k'+k}=\{S_j\}_{j=1}^m}\,
\sum_{\left(\Z^3\right)^{2l}}\,R_r(x,x_1)R_r(x,x_{l+2})\,\\
\times\ \prod_{j=1}^m c_{|S_j|}\delta(x_{S_j})\,
\prod_{i\in\Upsilon(\pi_{k'+k})}R_r(x_i,x_{i+1}) \
\hat{\prod_{j\in\Upsilon^c(\pi_{k'+k})}}R_r(x_{j+1},x_{j+2})\delta_{|x_{j+1}
- x_{j}|} \,.
\end{multline}
Therefore, the expectation of the rhs of (\ref{eq:re}) is given by
the formula
\begin{multline}
\E\, A^2_l(x,y)\ = \
\sum_{\beta=0}^l\,\lambda^{2l-2\beta}\,\sigma^\beta(E)\
\sum_{k,k':\{k+k'=\beta\}}\,(-1)^k\, \Big(_{\ \ \,k'\,\ }^{
\,k'+k\,}\Big)
\\ \times\  \sum_{m=1}^{l-\beta}  \, \sum_{\hat\pi_{\beta}=\{S_j\}_{j=1}^m}\,
\sum_{\left(\Z^3\right)^{2l}}\,R_r(x,x_1)R_r(x,x_{l+2})\\
\times\ \prod_{j=1}^m c_{|S_j|}\delta(x_{S_j})\,
\prod_{i\in\Upsilon(\pi_{\beta})}R_r(x_i,x_{i+1}) \
\hat{\prod_{j\in\Upsilon^c(\pi_{\beta})}}R_r(x_{j+1},x_{j+2})\delta_{|x_{j+1}
- x_{j}|} \,,
\end{multline}
and the only non vanishing contribution comes from $\beta=0$, since
\[\sum_{k,k':\{k+k'=\beta\}}\,(-1)^{k'}\,\, \Big(_{\ \ \,k'\,\ }^{
\,k'+k\,}\Big)\ = \ (1-1)^\beta\,,\] hence (\ref{eq:gaf1}).
\end{proof}

At this point we have to introduce additional notation (borrowed
from \cite{EY}):

\begin{defn} \label{def:equiv} We consider products of delta functions
with arguments that are linear combinations of the momenta $\{ p_1,
p_2,\ldots , p_{2 n +2}\}$. Two products of such delta functions
are called  {\em equivalent} if they determine the same affine
subspace of $\T^{2 n+2} = \{ p_1, p_2,\ldots , p_{2 n +2}\}$.
\end{defn}

One can obtain new delta functions from the given ones, by taking
linear combinations of their arguments. In particular, we can obtain
identifications of momenta.

\begin{defn}  \label{def:forced} The product of delta functions
$\Delta_\pi$  {\em forces} a new delta function $\delta (\sum_j a_j
p_j)$, if $\sum_j a_j p_j =0$ is an identity in the affine subspace
determined by $\Delta_\pi$.
\end{defn}

One can readily see that in the integrand of rhs of (\ref{eq:gaf})
one has a forced delta function
$\delta(p_1-p_{l+1}+p_{l+2}-p_{2l+2})$, hence
\begin{multline}\label{eq_an}
A_{n,E^*}(x-y):=\E \,A^2_n(x,y)\ = \ \lambda^{2n}\,\int
\,e^{-i2\pi(p_1+p_{n+2})\cdot (x-y)}\ \prod_{t=1}^{2n+2}dp_t
\,\frac{1}{e(p_t)+E^*}
\\ \times \ \sum_{m=1}^n\sum_{\pi=\{S_j\}_{j=1}^m}^{\ \ \
\ \ '} \prod_{j=1}^m c_{|S_j|}\delta(\sum_{i\in S_j}p_i-p_{i+1}) \,.
\end{multline}

\section{Proof of lemma \ref{lemma:rep}}\label{section:lemma}
Let us start with a remark that although the proof below is inspired
(and closely follows) by work of Erdos and Yau \cite{EY}, the
geometry of energy surfaces plays no significant role here. In
\cite{EY}, the geometry of energy surfaces poses a central difficult
problem because the energy parameter varies, and in particular
assumes values in the bulk of the essential spectrum of the nearest
neighbor Laplacian. However, in the situation discussed in this
paper, the reference energy $E^*$ is fixed, and lies below
$\inf(\sigma(-\Delta))=0$.
\subsection{Reduction to the pairing case}
In order to estimate $A_{n,E^*}(x-y)$ it suffices (up to the
combinatorial factors) to consider the special case of partitions
$\pi$ that appears in (\ref{eq:gaf1}), where $\pi=\{S_j\}_{j=1}^n$
with ${\rm card}\ S_j=2$ for each $j$ - the so called pairing case.
All other contributions are dominated by the corresponding pairing
counterparts, as becomes transparent from the positivity of the free
lattice Green function $R_r(x,y)$, cf. (\ref{eq:stand}). Indeed, for
any partition $\pi= \{S_j\}_{j=1}^m$ choose an arbitrary
subpartition $\pi'= \{S'_j\}_{j=1}^m$ into pairs. Evidently,
\begin{multline}\label{eq:redu}
\sum_{\left(\Z^3\right)^{2n}}\,R_r(x,x_1)R_r(x,x_{l+2})\,
\prod_{i\in\Upsilon_{l,l}}R_r(x_i,x_{i+1})\, \prod_{j=1}^m
\delta(x_{S_j}) \\ \le \
\sum_{\left(\Z^3\right)^{2n}}\,R_r(x,x_1)R_r(x,x_{l+2})\,
\prod_{i\in\Upsilon_{l,l}}R_r(x_i,x_{i+1})\, \prod_{j=1}^m
\delta(x_{S'_j}) \,,
\end{multline}
while the factor $\prod_{j=1}^m |c_{|S_j|}|$ is bounded by
$(cn)^{2n+1}$, see discussion in Subsection \ref{sub:prod}. Hence,
if one gets some bound $M$ on the pairing type contributions, the
whole $A_{n,E^*}(x-y)$ term can be rudely estimated as
$(2cn^2)^{2n+1}M$ (where we took into the account the number of the
possible partitions).

\subsection{Feynman graphs}\label{sub:F}

\begin{figure}[ht] \hskip -1 in
\includegraphics[width=11cm]{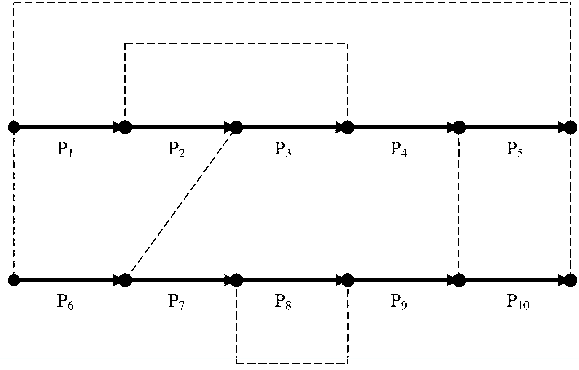}
\caption{Construction of the Feynman graph, part I, $n=4$. The
corresponding delta functions are $\delta(p_1-p_2+p_3-p_4)$,
$\delta(p_4-p_5+p_9-p_{10})$, $\delta(p_2-p_3+p_6-p_7)$, and
$\delta(p_7-p_9)$. The last delta corresponds to the tadpole. Note
that the sum of all momenta in the above delta functions gives a
forced delta function $\delta(p_1-p_5+p_6-p_{10})$, hence we can
introduce the dashed lines connecting vertices $1,6,7,$ and $12$,
identifying them as a single vertex.} \label{fig:con}
\end{figure}

\begin{figure}[ht]
\hskip -0.8 in
\includegraphics[width=10cm]{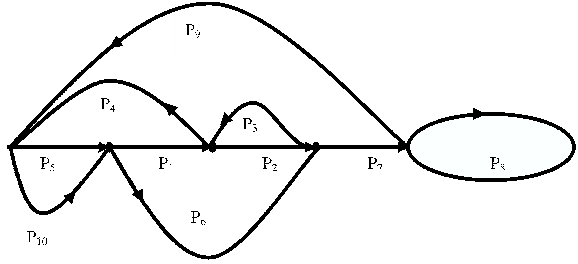}
\caption{Construction of the Feynman graph, part II: Identification
of the vertices. The tadpole corresponds here to $0$-loop.}
\label{fig:conII}
\end{figure}

$A_{n,E^*}(x-y)$ is conveniently  interpreted in terms of the so
called Feynman graphs (the pseudograph, to be precise, since loops
and multiple edges are allowed here). The graph, associated with
particular partition $\pi$ of $\Upsilon_{n,n}$ is constructed
according to the following rules (see Figure \ref{fig:con} and
\ref{fig:conII}): We first draw two line segments, each containing
$n$ vertices (elements of $\Upsilon_{n,n}$).
 The vertices are joined by directed edges (momentum lines) representing momenta:
$p_1, \ldots , p_{n+1}$ and $p_{n+2}, \ldots , p_{2n+2}$. To each
line $p_j$ we assign a propagator $F(p_j)$, with some given function
$F$, save momentum lines $p_1$ and $p_{n+2}$, which carry additional
phases $e^{-i2\pi p_1\cdot (x-y)}$ and $e^{-i2\pi p_{n+2}\cdot
(x-y)}$, respectively. For $\pi=\{S_j\}_{j=1}^m$ we identify all
vertices in each subset $S_j$ as the same vertex (in  Figure
\ref{fig:con}, the paired vertices are connected by  dashed lines).
\vspace{-.4cm}
Note that thanks to the existence of the forced delta function
$\delta(p_1-p_{l+1}+p_{l+2}-p_{2l+2})$, we can identify vertices
$1,n,n+1,2n$ as a single one, and therefore one can think about the
closed graph (with special rules that apply for momentum lines $p_1$
and $p_{n+2}$, mentioned above). To summarize, the outcome of this
construction is a directed closed graph, which is called the Feynman
graph associated with the partition $\pi$. The momenta in the graph
satisfy the Kirchhoff's first law, that is the total momenta
entering into each internal vertex add up to zero (if arrow faces
outward the vertex, we count its momentum with a minus sign). A
tadpole corresponds to the so-called {\em $0$-loop}, that is some
(directed) line of the graph claims one vertex as its both
endpoints. For a given Feynman graph $G$, one can choose a
particularly useful expression for the product of delta functions
$\Delta_\pi$. Choose any spanning tree of $G$ which does not contain
momentum lines $p_1,p_{n+2}$. The edges belonging to the spanning
tree will be called the {\em tree} edges (momentum lines), and all
the rest are the {\em loop} edges (since an addendum of any loop's
momentum line creates a loop). Let us enumerate the tree variables
as $u_1,...,u_k$, and loop variables as $w_1,...,w_l$, with say
$w_1=p_1,w_2=p_{n+2}$  (note that $k+l=2n+2$). The number $k$ of the
tree momenta coincides with the number of the delta functions in
$\Delta_\pi$. \vspace{-.4cm}
 One can check (see e.g. \cite{EY}) that the product of
delta functions $\Delta_\pi$ is equivalent to
\begin{equation}\label{eq:deltapr}
\prod_{i=1}^k \delta(u_i-\sum_{j=1}^la_{ij}w_j)\,,
\end{equation}
with
\[
a_{ij}:= \begin{cases} \pm1 & \hbox{ loop is created by adding $w_j$
to the spanning tree contains $u_i$}  \cr
 0
 &  \hbox{
otherwise} \cr
\end{cases}\,.
\]
The choice of the sign depends on the mutual orientation of $u_i$
and $w_j$.

\subsection{Exponential decay}

We first want to establish the exponential decay of $A_{n,E^*}(x)$
that appears (\ref{eq_an}) from the simple analytic argument, and
then in the next section obtain the bound on $A_{n,E^*}(0)$.

Specifically, we want to show that for a general value of $n$,
\begin{equation}\label{eq:expdec}A_{n,E^*}(x)\le
e^{-|x|\sqrt{E^*/3}}\,A_{n,E^*/2}(0)\,.\end{equation} Indeed, for
any given $x\in \Z^3$ lets choose $\gamma\in\{1,2,3\}$ such that
\begin{equation}\label{eq:ga}
|x\cdot e_\gamma|=\max_{i\in\{1,2,3\}} |x\cdot e_i|\,.
\end{equation}
Then $|x\cdot e_\gamma|\ge|x|/\sqrt3$. In order to obtain the
exponential decay of $A_{n,E^*}(x)$ we first perform the integration
in the rhs of (\ref{eq_an}) over the tree momenta, using
(\ref{eq:deltapr}). Let us use the shorthand notation $\sum_\pi$ for
a sum over all possible partitions in (\ref{eq:deltapr}), $c_\pi$
for a product of the corresponding $c_{Sj}$, and $r_\pi$ will denote
the number of the delta functions containing the loop  momentum
$w_1$ in the $\pi$'s partition. We get
\begin{multline}
A_{n,E^*}(x)\ = \ \lambda^{2n} \sum_\pi c_\pi\int
dw_1\prod_{i=1}^{r_\pi} \,\frac{1}{e(w_1+q_i)+E^*}\,e^{-i2\pi
w_1\cdot x}\, \cdot
\\ \int \prod_{t\in \Phi'}dp_t\,e^{-i2\pi w_{2}\cdot
x}\,\prod_{i=r_\pi+1}^{2n+2}\ \,\frac{1}{e(q_i)+E^*} \,,
\end{multline}
where $\Phi'$ is a set of all loop variables in the partition $\pi$,
except for $w_1$. The variable $q_k$ is some linear combination of
the loop variables in $\Phi'$, for $k=1,...,2n+2$. We want to obtain
a bound on the integral over $w_1$ variable. Note that
\begin{multline}
\int dp\prod_{i=1}^r \,\frac{1}{e(p+q_i)+E^*}\,e^{-i2\pi p\cdot x}\\
= \ \int dp' \,e^{-i2\pi (p\cdot x-p\cdot e_\gamma x\cdot
e_\gamma)}\int_{-1/2}^{1/2} d(p\cdot e_\gamma)\prod_{i=1}^r
\,\frac{1}{e(p+q_i)+E^*}\,e^{-i2\pi (p\cdot e_\gamma  x\cdot
e_\gamma)}\,,
\end{multline}
where $\int dp'$ stands for integration over components of $p$
orthogonal to $e_\gamma$. Without loss of generality, let us assume
that $x\cdot e_\gamma>0$. It is easy to check that the integrand as
a function of $p\cdot e_\gamma$ is $1$-periodic, analytic  inside
the rectangle formed by the points $$\{-1/2;\ -1/2+i\sqrt{E^*}/5;\
1/2+i\sqrt{E^*}/5;\ 1/2\}$$ for sufficiently small $E^*$: Indeed, we
have
$${\rm Re}\, e(p+q+i\epsilon e_\gamma )+E^*\ge e(p+q)+E^*/2$$
uniformly in $q$, provided
\[0\le\epsilon\le
\frac{\sqrt{E^*}}{3}\] for small $E^*$, where we have used
$\sin(a+ib)=\sin a \cosh b +i \cos a \sinh b$. Moreover, the
periodicity implies that the integrals over the vertical segments
coincide:
\begin{multline}
\int_{-1/2}^{-1/2+i\sqrt{E^*}/(2 \pi)} d(p \cdot
e_\gamma)\prod_{i=1}^r \,\frac{1}{e(p+q_i)+E^*}\,e^{-i2\pi (p\cdot
e_\gamma  x\cdot e_\gamma)}
\\ = \
\int_{1/2}^{1/2+i\sqrt{E^*}/(2 \pi)} d(p\cdot e_\gamma)\prod_{i=1}^r
\,\frac{1}{e(p+q_i)+E^*}\,e^{-i2\pi (p\cdot e_\gamma  x\cdot
e_\gamma)}\,.
\end{multline}
Therefore
\begin{multline}
\left|\int_{-1/2}^{1/2} d(p\cdot e_\gamma)\prod_{i=1}^r
\,\frac{1}{e(p+q_i)+E^*}\,e^{-i2\pi (p\cdot e_\gamma  x\cdot
e_\gamma)}\right| \\
= \ \left|\int_{-1/2+i\sqrt{E^*}/(2 \pi)}^{1/2+i\sqrt{E^*}/(2 \pi)}
d(p\cdot e_\gamma)\prod_{i=1}^r \,\frac{1}{e(p+q_i)+E^*}\,e^{-i2\pi
(p \cdot e_\gamma
x\cdot e_\gamma)}\right|\\
\le \ e^{-x\cdot e_\gamma\sqrt{E^*}}\int_{-1/2}^{1/2}
d(p\cdot e_\gamma)\prod_{i=1}^r \,\frac{1}{e(p+q_i)+E^*/2}\\
\le \ e^{-|x|\sqrt{E^*/3}}\,\int_{-1/2}^{1/2} d(p\cdot
e_\gamma)\prod_{i=1}^r \,\frac{1}{e(p+q_i)+E^*/2}\,,
\end{multline}hence (\ref{eq:expdec}).


\subsection{Bound on the pairing type diagrams}
Note that for any $ p\in \T^3$ we have an elementary estimate
$$e(p)\ =\ 2\sum_{i=1}^3\sin^2(\pi p\cdot e_i)
 \ > \ 2 \ \sin^2\Big(\frac{\pi |p|}{\sqrt3}\Big)\,.$$ On the other
hand, by Jordan's inequality $\sin^2\Big(\frac{\pi
|p|}{\sqrt3}\Big)\ge \frac{4p^2}{3}$ for any ${ p}\in \T^3$.
Therefore,
\begin{equation}\label{eq:quad}(e({ p})+E^*)^{-1}\ \le \
C(p^2+E^*)^{-1}\end{equation} for any ${ p}\in \T^3$, and
\begin{multline}\label{eq_an1}
A_{n,E^*}(0)\ \le \ \lambda^{2n} \int \ \prod_{t=1}^{2n+2}dp_t
\,\frac{C}{p^2_t+E^*}
\\ \times \ \sum_{m=1}^n\sum_{\pi=\{S_j\}_{j=1}^m}^{\ \ \
\ \ '} \left(\prod_{j=1}^m |c_{|S_j|}|\right)\delta(\sum_{i\in
S_j}p_i-p_{i+1}) \,.
\end{multline} Now consider the pairing type partition $\pi$. For
such a partition all $c_{|S_j|}=1$.  It is natural to rescale
variables, $p_t=\sqrt{E^*}q_t$, to get
\begin{multline}\label{eq_an2}
\int \ \prod_{t=1}^{2n+2}dp_t \,\frac{C}{p^2_t+E^*}  \
\delta(\sum_{i\in S_j}p_i-p_{i+1}) \\ = \
(E^*)^{-\frac{n}{2}+1}\,\int^{\ '} \ \prod_{t=1}^{2n+2}dq_t
\,\frac{C}{q^2_t+1}  \ \prod_{j=1}^n \ \delta(\sum_{i\in
S_j}q_i-q_{i+1}) \,,
\end{multline}
where we used the scaling $\delta(a { p})=a^{-3}\delta({ p})$ in
three dimensions (and the fact that there are $n$ delta functions
involved in the pairing case). Let us note that $n$ is equal to the
number of loop momenta and equal to the number of vertices minus
$1$. Each integration now runs over the torus $(E^*)^{-1/2}\T^3$,
concisely denoted as $\int^{\ '}$. We can only increase the right
hand side by replacing it with
\begin{multline}\label{eq_an3}
C^{2n+2}\,(E^*)^{-\frac{n}{2}+2}\,\int^{\ '} \
\prod_{t=1}^{2n+2}dq_t
\,\frac{\ln^4((E^*)^{-1}+1)}{(q^2_t+1)\ln^4(q^2_t+2)}  \
\prod_{j=1}^n \ \delta(\sum_{i\in S_j}q_i-q_{i+1}) \\ < \ \ \tilde
C^{2n+2}\,\frac{\ln^{8n+8}E^*}{(E^*)^{\frac{n}{2}-1}}
\int_{(\R^3)^{2n+2}} \ \prod_{t=1}^{2n+2}dq_t
\,\frac{1}{(q^2_t+1)\ln^4(q^2_t+2)} \ \prod_{j=1}^n \
\delta(\sum_{i\in S_j}q_i-q_{i+1})
 \,,
\end{multline}
where $\delta(\cdot)$ in the last line stands for the standard Dirac
delta distribution on $\R^3$ .

At this point we are going to switch to the (Euclidean) Feynman
graph representation, as described in Subsection \ref{sub:F}, with
 the propagator
\[F(q_j):=\frac{1}{(q^2_j+1)\ln^4(q^2_j+2)}\,.\]

Let us define the {\em value} $|G|$ of the graph $G$ (characterized
by its partition $\pi=\{S_j\}_{j=1}^n$) as
\begin{equation}\label{eq:pai}
|G|=\int_{(\R^3)^{2n+2}} \ \prod_{t=1}^{2n+2}dq_t
\,\frac{1}{(q^2_t+1)\ln^4(q^2_t+2)} \  \
 \prod_{j=1}^n \
\delta(\sum_{i\in S_j}q_i-q_{i+1}) \,.
\end{equation}

If partition $\pi$ is tadpole-free (and those are only partitions
that enter into (\ref{eq_an}) thanks to the self energy
renormalization), the corresponding Feynman diagram has an important
property of being (logarithmically) superficially convergent. The
value of such graphs is controlled by the following theorem (an
adaptation of Theorem $A1$ in \cite{FMRS} to the case in hand):

\begin{thm}[Bound on superficially convergent Euclidean Feynman graphs,
\cite{FMRS}]\label{thm:super} If  Feynman graph satisfies
assumptions ${\mathcal A}1$-${\mathcal A}3$ below, then its value is
bounded by $K^{n}$, with some generic constant $K$.
\end{thm}
This result yields that the contribution from a pairing type
partition to $A_{n,E^*}(0)$ is bounded by
\begin{equation}\label{eq:pai1}E^*\ \left(C\ \ln^{9}E^*\
\frac{\lambda^{2}}{\sqrt{E^*}}\right)^n\end{equation} - the key
estimate of this subsection.
\\
The subsequent assumptions require the introduction of some
additional notation. Let a {\it subgraph} $G'\subseteq G$ be a
subset of the lines of $G$. The vertices of $G'$ are the end points
of lines of $G'$ and an external line of $G'$ is an edge of
$G\setminus G'$ which is hooked to a vertex of $G'$.

Let us denote by $ \Lambda(G')$ the number of loop edges of $G'$
(defined in Subsection \ref{sub:F}), by $E(G')$ the number of
external edges of $G'$, by $I(G')$ the number of internal lines of
$G'$, and by $N(G')$ the number of vertices in $G'$.

Two subgraphs $F$, $F'$ are {\it disjoint} if they have no line and
no vertex in common. They {\it overlap} if they are not disjoint and
do not satisfy an inclusion relation ($F\subseteq F'$ or
$F'\subseteq F$). A {\it forest} $\mathbb F$ is a set of
non-overlapping connected subgraphs.

A subgraphs $F$ is said to be {\it one line reducible} (OLR) if
there exists a line of $F$ such that its removal increases the
number of connected components of $F$. A connected subgraph which is
not OLR is called {\it proper}.

\begin{enumerate}
\item[${\mathcal A}1$] {\it The lines}\\
Each line has a  propagator of the form
\[\frac{1}{(q^2+1)\ln^4(q^2+2)}\,.\]

\item[${\mathcal A}2$] {\it Superficial convergence}\\
For any connected subgraph $G'\subseteq G$ a superficial degree of
divergence ${div}(G')$ is given by
\[{  div}(G) \ = \ 3\Lambda(G)\ - \ 2I(G)\]

For any connected subgraph $G'\subseteq G$ a logarithmical
superficial degree of divergence ${l-div}(G')$ is given by
\[{l-div}(G') \ = \ \Lambda(G')\ - \ 4I(G')\,.\]
A graph $G$ is called {\em superficially convergent} if any
connected subgraph $G'$ of $G$ satisfies either
\[{ div}(G')< -2\epsilon E(G')\]or
\[div(G')=0\ {\rm and}\ l-div(G')\le-\epsilon\,.\]
\item[${\mathcal A}3$] {\it $div=0$ Forests}\\
(i)\ \ There exists a constant $C$ such that the number of proper
$div=0$ forests of $G$ (i.e. forests consisting of proper subgraphs
$G'$ with $div(G')=0$) is bounded by $C^{L(G)}$.
\\
(ii)\ \ There exists a constant $C'$ such that for every connected
$G'\subseteq G$  with $div(G')=0$, $G'$ has at most $C'$ external
vertices.
\end{enumerate}

\begin{remark}
Theorem $A1$ in \cite{FMRS} is a much more general result than the
one presented here. It contains two additional assumptions which are
irrelevant in our context (namely $HA.2-HA.3$ in \cite{FMRS}). We
also adapted the various definitions from \cite{FMRS} (such as a
superficial degree of divergence) to the concrete situation
discussed in this paper.
\end{remark}

\begin{figure}[ht] \hskip -1.5 in
\includegraphics[width=7cm]{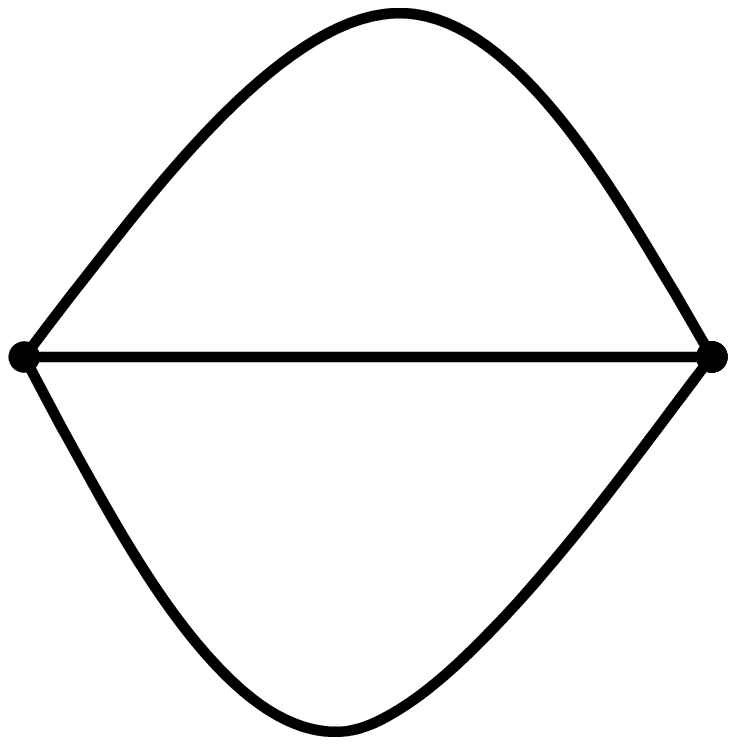}\vskip -7.6cm
\caption{Graph $F$.} \label{fig:0}
\end{figure}

The rest of this subsection is devoted to the validation of the
assumptions ${\mathcal A}2$-${\mathcal A}3$ of Theorem
\ref{thm:super} in the present context. ${\mathcal A}3$ is met since
all superficially divergent subgraphs in our situation (in fact
there is only one divergent subgraph $F$, introduced below on Fig.
\ref{fig:0}) are also divergent in $\phi_4^4$ theory, where this
assumption holds true with $C=8$ (\cite{dCR}). The constant $C'$
(which in our context corresponds to the number of external lines of
the aforementioned graph $F$) is equal to $2$.

We now want to establish the validity of ${\mathcal A}2$, with
$\epsilon=\frac{1}{10}$:

Let us note that for a pairing partition, the degree of each
internal vertex is $4$, that is we have $4$-regular directed graph.
For such a graph any spanning tree contains $n$ edges, and $n+2$
loops accordingly.

Since $G$ is $4$-regular, it is easy to see that for any connected
subgraph $G'$
\begin{equation}\label{eq:cou}
I(G')\le \frac{4N(G')-E(G')}{2}\,;\quad \Lambda(G')+N(G')-1=I(G')\,,
\end{equation}
where the latter relation follows from the fact that the spanning
tree for $G'$ contains $N(G')-1$ lines, and the rest of the internal
lines can be thought of as loops. Hence
\begin{multline}\label{eq:coun}
{  div}(G')  \ = \ -\Lambda(G')+2(2\Lambda(G')-I(G')) \\ = \
-\Lambda(G')+2(I(G')+2-2N(G')) \ \le \ 4-E(G')-\Lambda(G')
\end{multline}
and
\begin{multline}\label{eq:coun11}
{ l-div}(G')  \ = \ \left(\Lambda(G')-I(G')\right)-3 I(G')\ = \
1-N(G'))-3 I(G') \ \le \ -4
\end{multline}
for  any subgraph $G'$ of $G$ with $N(G')\ge2$. Since $N(G')=1$
corresponds to the tadpole, which is not allowed, we  conclude than
is that all relevant subgraphs are logarithmically convergent.

For the whole graph $G$ we have $div (G)=3(n+2)-2(n+2)<-n/4$ for
$n>2$, while $div(G)=0$ for $n=2$. Since $2 E(G') \mod 4=0$ for any
subgraph $G'$ of the $4$-regular graph $G$, and $E(G')\neq0$ unless
$G'=G$, we deduce from (\ref{eq:cou}) and (\ref{eq:coun})  that for
any Feynman graph $G$ corresponding to (\ref{eq_an3}) with $n\ge2$
the only possible proper subgraphs $G'$ with $div (G')\ge0$ are:
\begin{enumerate}
\item $N(G')=1,\ \Lambda(G')=1,\ E(G')=2$ - a $0$-loop, that is a
tadpole. For the $0$-loop $div(G')=1$, but on the other hand, the
tadpoles are prohibited in our partition.
\item $N(G')=2,\ \Lambda(G')=2,\ E(G')=2$ - either a pair of
tadpoles connected by an edge or a graph $F$, shown on Figure
\ref{fig:0} (with two external edges omitted). For the latter graph
we have $div(F)=0$, hence $F$ is superficially convergent as well.
\end{enumerate}
\vskip -.6cm
Since it is clear from (\ref{eq:coun}) that ${
div}(G')\le -E(G')/5$ for $E(G')\ge5$ while ${  div}(G')\le
-1\le-E(G')/5$ for $E(G')\le5$ (with the exception of the graph
$F$), we conclude that all tadpole-free pairings are indeed
superficially convergent.


\subsection{Bounds (\ref{eq:l}--\ref{eq:a1})}

We deduce from (\ref{eq:pai1}), (\ref{eq:redu}) and discussion
thereafter that
\[A_{n,E^*}(0)\ \le \ (4n)!\,
E^*\,\left(C\ \ln^{9}E^*\
\frac{\lambda^{2}}{\sqrt{E^*}}\right)^n\,,\] and (\ref{eq:l})
follows now from (\ref{eq:expdec}).

To get (\ref{eq:lt}) note that it follows from Lemma \ref{lem:de}
that
\[\tilde A_N \ = \ A_N(-\frac{1}{2}\Delta+E^*) \ - \ \sigma(E)
A_{N-1}R_r\,.\] We therefore obtain
\begin{multline}
\E \,|\tilde A_N(x,y)| \ \le \ \sum_{z\in\Z^3}\left\{\,\left(\E
\,|A_N(x,z)|^2\right)^{1/2}\,\cdot\,
|(-\frac{1}{2}\Delta+E^*)(z,y)|\right.
\\ \left.
+ \ \sigma(E)\, \left(\E
\,|A_{N-1}(x,z)|^2\right)^{1/2}\,\cdot\,R_r(z,y)\,\right\}
\\ < \
7\sum_{z\in\Z^3:\ |z-y|\le 2}\left(\E
\,|A_N(x,z)|^2\right)^{1/2} \\
+ \ C\sum_{z\in\Z^3} \sigma(E)\, \left(\E
\,|A_{N-1}(x,z)|^2\right)^{1/2}\,\cdot\,\frac{1}{|z-y|+1}\,e^{-\sqrt{2E^*}|z-y|}
\,,\nonumber
\end{multline}
with some generic constant $C$, provided $E^*\le1$ and where we used
the bound (\ref{eq:asymp}) on the free Green function $R_r(z,y)$.

It is clear from (\ref{eq:l})  that the first contribution is
bounded by
\[C'\,\sqrt{(4N)!\,E^*}
\,\left(C\ \ln^{9}E^*\ \frac{\lambda^{2}}{\sqrt{E^*}}\right)^{N/2}
\,e^{-\sqrt\frac{{E^*}}{12}\,|x-y|}\,,\] while the second one is
bounded by
\begin{multline}
C'\,P(\lambda,E^*)
\sum_{z\in\Z^3}\frac{1}{|z-y|+1}\,e^{-\sqrt\frac{{E^*}}{12}\,|x-z|}\,e^{-\sqrt{2E^*}|z-y|}
\\ < \ C'\,P(\lambda,E^*)\,e^{-\sqrt\frac{{E^*}}{12}\,|x-y|}\,
\sum_{z\in\Z^3}\frac{1}{|z-y|+1}\,e^{-\sqrt{2E^*}(1-1/\sqrt{24})|z-y|}
\\ < \ \frac{\tilde C}{E^*}\,P(\lambda,E^*)\,e^{-\sqrt\frac{{E^*}}{12}\,|x-y|}
\,,\nonumber
\end{multline}
with \[P(\lambda,E^*):=\lambda^2\,\sqrt{(4N-4)!\,E^*} \,\left(C\
\ln^{9}E^*\ \frac{\lambda^{2}}{\sqrt{E^*}}\right)^{N/2-1}\,,\] hence
(\ref{eq:lt}).

The positivity of $A_0(x,y)$ in (\ref{eq:stand}) is immediate from
the positivity of $e^{t\Delta}(x,y)$ for all non negative values of
$t$, which in turn is obvious, since all off diagonal matrix
elements of the Laplacian are positive, and its diagonal part is
proportional to the unity operator.

The upper bound in (\ref{eq:stand}) is the standard estimate, see
e.g. \cite{Lawler}, it also follows from (\ref{eq:asymp}). The bound
(\ref{eq:asymp}) for the free Green's function is known as well, see
e.g. \cite{KI}.

The convenient way to establish bound (\ref{eq:a1}) is in the
coordinate representation: Since $\E \,(V_\omega(x)
V_\omega(y))=\delta_{|x-y|}$ one checks that
\[\E\, A_1^2(x,y)=\lambda^2
\,\sum_{z\in\Z^3}R^2_r(x,z)R^2_r(z,y)\,.\] Now we use
(\ref{eq:asymp}) to bound the rhs from above  as
\begin{multline}
 \lambda^2\,\sum_{z\in\Z^3}\frac{1}{ |x-z|^2+1}\,\frac{1}{
|y-z|^2+1}\,e^{-2\sqrt{2E^*}(|x-z|+|y-z|)} \\ \le \
\lambda^2\,e^{-2\sqrt{2E^*}|x-y|}\,\sum_{z\in\Z^3}\frac{1}{
|x-z|^2+1}\,\frac{1}{ |y-z|^2+1}\\
\le \ \frac{ C\, \lambda^2}{|x-y|+1}\,e^{-2\sqrt{2E^*}|x-y|} \,.
\end{multline}


\appendix

\section{Properties of the solution of
(\ref{eq:self})}\label{sec:appendI} It is instructive to rewrite
(\ref{eq:self}) as
\begin{equation}\label{eq:self1}
E=E^*+\lambda^2\int_{\T^3} \frac{d^3p}{e(p)+E^*}\,.
\end{equation}
$E$ then is a well defined function of $E^*$ on the ray
$E^*\in[0,\infty)$, with \[E(0)=\lambda^2\int_{\T^3}
\frac{d^3p}{e(p)}\,.\] Differentiation of (\ref{eq:self1}) with
respect to $E^*$ gives
\begin{equation}\label{eq:self2}
E'(E^*)=1-\lambda^2\int_{\T^3} \frac{d^3p}{(e(p)+E^*)^2}\,.
\end{equation}
Monotonicity of the integral on the rhs of the above equation for
$E^*\in(0,\infty)$ implies that there exists exactly one extremum of
$E$ (namely minimum) for such values of $E^*$. In particular, the
function $E=E(E^*)$ is invertible for $E>E(0)$. For any $E^*>0$ we
have an estimate
\[\int_{\T^3} \frac{d^3p}{(e(p)+E^*)^2}\le C(E^*)^{-1/2}\,,\]
which follows from (\ref{eq:quad}) and the extension of the domain
of integration to the whole $\R^3$. We infer that
\[E(E^*)=\int_0^{E^*}E'(t)dt+E(0)\ge \lambda^2\int_{\T^3} \frac{d^3p}{e(p)}+E^*-C\lambda^2
\int_0^{E^*}t^{-1/2}dt\,,\] hence $E(E^*)> E(0)$ for $E^*>
4C^2\lambda^4$, so there exists an inverse function $E^*=E^*(E)$ in
this range. On the other hand, note that $e(p)\le p^2/2$, hence
\[\int_{\T^3} \frac{d^3p}{(e(p)+E^*)^2}\ge\int_{\T^3} \frac{d^3p}{(p^2+E^*)^2}
\ge\int_{B_{E^*}} \frac{d^3p}{(p^2+E^*)^2}=C' (E^*)^{-1/2}\,,\]
where $B_{E^*}$ denotes the ball of radius $\sqrt{E^*}$ around the
origin ($E^*$ is assumed to be small). Therefore we obtain
\[E\le\lambda^2\int_{\T^3} \frac{d^3p}{e(p)}+E^*-C'\lambda^2
\int_0^{E^*}t^{-1/2}dt\,,\] so that if
\[E\ge \lambda^2\int_{\T^3}
\frac{d^3p}{e(p)} +\lambda^{4-\epsilon}\,,\] then $E^*\ge \tilde
C\lambda^{4-\epsilon}$ for small enough values of $\lambda$, as
follows from the solution of the corresponding quadratic equation
for $\sqrt{E^*}$.


 \bigskip

 \section*{Acknowledgements}
This contribution would not have been possible without all I learnt
from  H.-T. Yau to whom I am indebted. I also would like to thank L.
Erdos for a number of valuable discussions and suggestions. I am
grateful to the referees for many useful remarks and numerous
corrections.

\end{document}